\documentstyle[prl,aps,multicol,epsf]{revtex}
\renewcommand{\narrowtext}{\begin{multicols}{2}
\global\columnwidth20.5pc} 
\renewcommand{\widetext}{\end{multicols}
\global\columnwidth42.5pc} \multicolsep = 8pt plus 4pt minus 3pt

\newcommand{\be}{\begin{equation}}
\newcommand{\ee}{\end{equation}}

\begin{document}
\draft
\title{
Transition from quantum Hall to compressible states in the
second Landau level:\\ new light on the $\nu$=5/2 enigma
}
\author{
R.H. Morf
}
\address{
Paul Scherrer Institute, Villigen, Switzerland
}
\date{7 October 1997}
\maketitle

\begin{abstract}
Quantum Hall states at filling fraction $\nu$=5/2 are examined by numerical
diagonalization.
Spin-polarized and -unpolarized states of systems with $N$$\le$18 electrons
are studied, neglecting effects of Landau level mixing.
We find that the ground state is spin polarized. It is incompressible and
has a large overlap with paired states like the Pfaffian.
For a given sample, the energy gap is about 11 times smaller than at $\nu$=1/3.
Evidence is presented of phase transitions to compressible states,
driven by the interaction strength at short distance.
A reinterpretation of experiments is suggested.
\end{abstract}

\narrowtext

Ten years after the discovery of a quantized Hall plateau at filling fraction
$\nu$=5/2 by Willett et al. \cite{Willett87}, `a key piece of the $\nu$=5/2
puzzle is still missing': This is the conclusion reached by Eisenstein
in his recent review \cite{perspectives,Eisenstein_book}.
Studies by Eisenstein et al. \cite{Eisenstein88} in a tilted magnetic field
had shown that
the plateau disappears when the tilt angle exceeds a critical value.
It is now widely believed that the plateau is the result of a
{\it spin-unpolarized incompressible} ground state (GS), while, at larger tilt
angles, the Zeeman energy favors a {\it polarized compressible} GS,
consistent with the disappearance of the plateau.

The evidence supporting the above picture is taken from activation studies 
which reveal an energy gap that decreases with increasing tilt angle
\cite{Eisenstein90}.
This fact is explained naturally if the GS is unpolarized
and if its lowest energy excitations involve electrons with reversed spin,
and thus a gain in Zeeman energy $\Delta$$E$=$g$$\mu_B$$B$ from spin-reversal
($g$ and $\mu_B$ stand for the $g$-factor and the Bohr magneton). This
energy gain increases with increasing tilt angle $\Theta$ as the magnetic field
perpendicular to the sample, $B_{\bot}$=$B$$\cos$$\Theta$, is fixed by the
electron density $n_S$ of the sample
and the filling fraction $\nu$ \cite{Eisenstein89}.
From the slope of the activation energy as a function of $B$, a $g$-factor
$g$$\approx$0.56 was extracted \cite{Eisenstein90,Eisenstein_book},
somewhat larger than its value $g$=0.44 for bulk GaAs.
That the polarized state expected at large tilt angles
should be compressible, is consistent with the
Fermion Chern-Simons theory of Halperin, Lee and Read \cite{HLR},
which predicts that electrons in a half-filled Landau level (LL) behave like
quasi-particles in zero magnetic field forming a Fermi liquid,
the `Composite Fermion (CF) liquid' \cite{Stormer_book}.

In this note, we challenge this interpretation of the experiments.
We present evidence from exact diagonalization results that the GS
in a half-filled second LL is spin-polarized and incompressible,
consistent with the prediction by d'Ambrumenil and the author
\cite{Morf95} that the CF-liquid does not form at this filling.

What makes the plateau disappear at large tilt angles?
If the system is spin-polarized already at small tilt angles, 
the Zeeman energy cannot drive the phase transition.
In this note, we show that the incompressible state is very sensitive to
details of the interaction: phase transitions to gapless states occur
when the interaction at short distance is either `too hard' or `too soft'.
When it is `too hard', we recover the compressible CF-liquid as GS.
We maintain that the system becomes gapless due to a phase transition
to a compressible state, driven by tilting the magnetic field, thereby
modifying the interaction.

In the following, we examine both spin-polarized and -unpolarized systems
by exact diagonalization \cite{partial}. We employ
Haldane's spherical geometry \cite{Haldane83}, in which
quantized Hall states at filling fraction $\nu_n$ of the $n$-th LL are
characterized by a specific relation between
the number of electrons $N$ and the number of flux quanta $N_\Phi$
\begin{equation}
N_\Phi=\nu_n^{-1} N - S.
\label{eq:nphi}
\end{equation}
Here, the `shift' $S$ depends on $\nu_n$ and the character of the FQH-state
\cite{dAmb}, and represents a topological quantum number \cite{Wen92}.
The value of $S$ for the FQH-state at $\nu$=5/2=2+1/2, i.e. $\nu_1$=1/2, is
not known although their exist definite predictions \cite{HR,HRprime,Moore}.
To locate the FQH-state, we make an unbiased
study for a whole range of $S$-values. We neglect LL mixing and approximate
the electron interaction by the Coulomb interaction of point
particles \cite{Morf_new}. As usual, the interaction is fully specified
by the values of Haldane's pseudopotentials \cite{Haldane83} $V_L$, i.e.
the interaction energy of two electrons with relative angular momentum $L$,
which is LL dependent.

Results of our exact diagonalizations are shown
in Fig. 1 \cite{secondLL}.
Energies per electron $E/N$ for spin-polarized and -unpolarized systems at
$\nu_1$=1/2, i.e.  $N_\Phi$=2$N$-$S$, are shown for different $S$. Fig. 1a
shows results for unpolarized, Fig. 1b those for polarized systems. In Fig. 1c,
we show the difference between GS-energies of unpolarized and polarized states.
Energies are quoted in units $e^2/\ell_0$ where $\ell_0$ denotes the magnetic
length, $\ell_0=\sqrt{\hbar c/eB}$, cf. \cite{Morf86}.
For even values of the flux $N_\Phi$, the GS
of the unpolarized systems has angular momentum $L$=0 in all
cases studied, cf. Fig. 1a. As incompressible states must be rotation-invariant
in the spherical geometry, they are, at least in principle, candidates for
FQH-states. Yet, they
\begin{figure}
\center
\epsfxsize=3.2truein
\hskip 0.0truein \epsffile{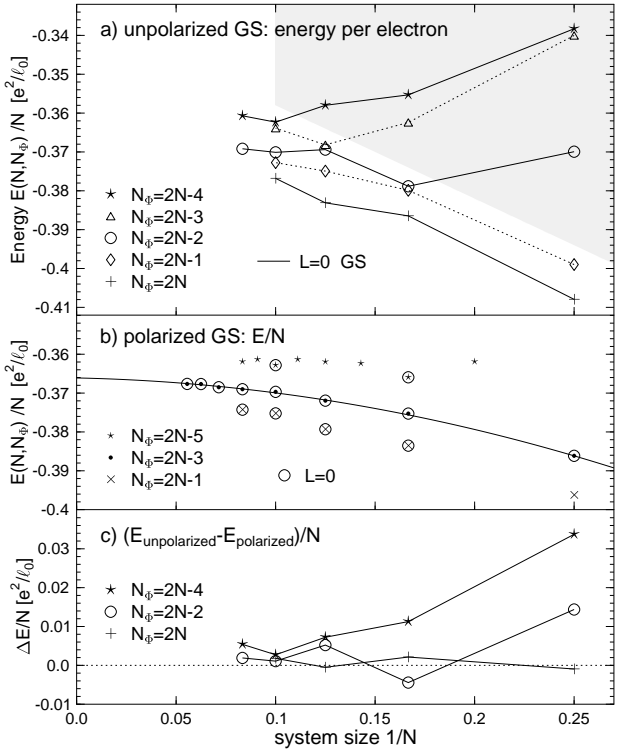}
\vspace{0.1in}
\caption{
For systems with 4$\le$$N$$\le$18 electrons and flux
$N_\Phi$=2$N$-$S$ GS energies $E/N$ are shown for unpolarized and polarized
systems.
a) Unpolarized system: $E/N$ for $S$=0,1,2,3,4. At even $N_\Phi$, GS
have angular momentum $L$=0. Results in the shaded area suffer from strong
finite size effects, cf. text.
b) $E/N$ at $S$=1,3,5 for polarized systems. At $S$=3 and even $N$, i.e. the
quantum numbers of the Pfaffian, all GS have angular momentum $L$=0.
c) Energy difference between unpolarized and polarized state for the same
$N,N_\Phi$.
}
\label{fig:fig1}
\end{figure}
\noindent
have in almost all cases
higher energy than polarized states at the same $N,N_\Phi$, cf. Fig. 1c.
The exception is the unpolarized state at $N$=6,
$N_\Phi$=10 (see Fig. 1c). At first, before investigating larger systems,
we were hopeful, that this observation might help to explain the
$\nu$=5/2 Hall plateau.
Our larger system studies do not support this hope:
For systems with up to $N$=12 electrons, no similar unpolarized state exists
and there is no hint that in the bulk limit,
the GS would be unpolarized \cite{LL_mixing}.

In fact, there is evidence that the properties of the GS at
$N$=6, $N_\Phi$=10 are not related to $\nu$=5/2:
Similar `cusps' in $E/N$ occur at $N$=8, $N_\Phi$=13 and
$N$=10, $N_\Phi$=16, cf. Fig. 1a. These appear on the line $N_\Phi$=3$N$/2+1,
which extrapolates to $\nu_1$=2/3 for large $N$, and have nothing to do
with the behavior at $\nu_1$=1/2. We believe that the cusps reflect a property
that for values of $N_\Phi$ below this line (corresponding to the
shaded area of Fig. 1a) it seems to be impossible to construct a
spin-singlet wf for which the pair correlation function $g(R)$ vanishes at
$R$=0, whereas it is possible to do so for $N_\Phi$ on or above
this line \cite{Hard_core}.
At filling $\nu_1$=1/2,  $N_\Phi$=2$N$-$S$ exceeds this limit
for large enough $N$. For smaller $N$, the GS have anoma-
\begin{figure}[t]
\center
\epsfxsize=3.2truein
\hskip 0.0truein \epsffile{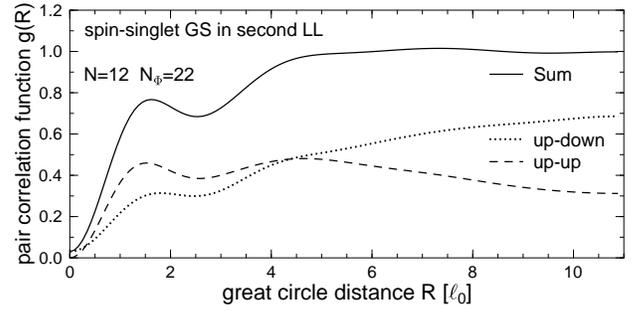}
\vspace{0.1in}
\caption{
Pair correlation function $g(R)$ of unpolarized GS for
$N$=12, $N_\Phi$=22 in the second LL.
}
\label{fig:fig2}
\end{figure}
\noindent
lously high energy, cf. shaded area in Fig. 1a.

To complement this picture, we show in Fig. 2 the pair correlation function
$g(R)$ for the unpolarized GS at $N$=12, $N_\Phi$=22,
and its components $g_{up-down}(R)$ and $g_{up-up}(R)$,
for electrons with unlike and like spins, respectively.
Clearly, $g(R)$ is close to zero as $R$ tends to zero.
In an unpolarized state, the number of electrons contributing
to $g_{up-down}(R)$ is $N/2$ while for $g_{up-up}(R)$ it is one less.
In a {\it local} spin-singlet, this extra electron is
close to the electron at the origin \cite{HR}, while in our state,
it is as far away as possible on a sphere, see Fig. 2. A system whose GS
is polarized, but in which a long-wavelength spin-excitation establishes
spin-singlet symmetry, would show such behavior.

In Fig. 1b, the GS energy $E/N$ of polarized states
is shown for systems with $N_\Phi$=2$N$-$S$, at $S$=1,3,5.
GS with $L$=0 are marked with circles. For even $S$,
GS have typically $L$$>$0 and are not shown.
Only for $S$=3, all GS for even $N$ are rotation invariant and
thus candidates for FQH states. Their energy increases smoothly with
size,  extrapolating to a bulk limit of $\approx$ -0.366.
Their flux agrees with predictions based on pair
formation \cite{HRprime,Moore}.

In Fig. 3a, we show the energy spectrum of a $N=8$ electron system for different
pair interactions, by varying the coupling strength $V_1$ in the $L_{rel}$=1
relative angular momentum channel, but keeping all the other $V_L$ at their
values for Coulomb interaction in the second LL. As we can see,
around $V_1$=1 (in units of $V_1^{Coulomb}$) there is a gap $\Delta$ in the
excitation spectrum $\Delta$$\approx$0.02. However, both for small and
large $V_1$, the gap disappears.

In Fig. 3b, we
show the overlap of the GS with the Pfaffian wf. Clearly, the overlap is
close to unity when $V_1$ has the value for Coulomb interaction. In fact,
overlap and gap have their maxima roughly at the same value $V_1$$\approx$1.1.
These results are consistent with conjectures by Greiter et al.
\cite{Greiter9192} that the $\nu$=5/2 FQH state might be related to the
Pfaffian.  However, this observation should not be overstated: Indeed, the GS
has a similarly large overlap with a pair wf \cite{Morf86}, setting
parameters $m$=1, $t$=0, $s$=2 in eq.(1)
of ref. \cite{Morf86}. In view of the ambiguity of trial
wf's, we cannot be sure that in the bulk limit, the GS  will exhibit the
characteristics of the Pfaffian, e.g.
\begin{figure}
\center
\epsfxsize=3.2truein
\hskip 0.0truein \epsffile{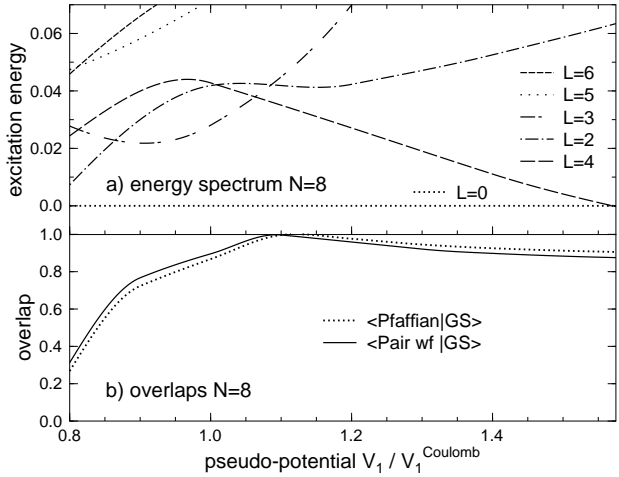}
\vspace{0.1in}
\caption{
a) Energy of low-lying states of polarized system of 8 electrons
in second LL at flux  $N_{\Phi}$=13 vs.
the $L$=1 pseudo-potential $V_1$, measured in units
$V_1^{Coulomb}$. The state becomes
gapless for small and large $V_1$.
b) Overlap of the GS wf with Pfaffian and pair-wf trial-states.
Gap and overlap both have their maximum at $V_1$$\approx$1.1 .
}
\label{fig:fig3}
\end{figure}
\noindent
excitations with non-Abelian statistics \cite{Moore,RR96}.

In Fig. 4a, we show the
excitation spectrum for a much larger system, $N$=16,$N_\Phi$=29.
The spectrum looks similar with a gap that vanishes
when $V_1$ is below 0.9 or larger than 1.3.
For Coulomb interaction, the gap is again $\Delta$$\approx$0.02
and its maximum still occurs at $V_1$$\approx$1.1. Similar
excitation spectra are also seen for sizes $N=$10 and 14, while the
system with $N$=12, $N_\Phi$=21 is `aliased' \cite{dAmb}
with a $\nu_1$=3/5 state  and its interpretation as a
$\nu_1$=1/2-state is dubious. The evidence for phase transitions to
gapless states for small and large $V_1$ appears firm.

The compressible state at large $V_1$ is the CF-liquid \cite{RR94,Morf95}. This
becomes clear from Fig. 4b. At $N$=16, the CF-state occurs at $N_\Phi$=30, one
flux unit higher than for the FQH state. As a reference CF-liquid wf, we use
the GS for Coulomb interaction in the lowest LL \cite{Morf95}.
As $V_1$ is increased, its overlap with the GS approaches
unity when the system becomes gapless. As
incompressible and CF-states do not exist at the same flux $N_\Phi$, a bias
exists in favor of the FQH state at $N_\Phi$=29,
whereas the CF-liquid is favored at $N_\Phi$=30.
The critical $V_1$ value will thus be either over- or underestimated,
depending on $N_\Phi$.

In Fig. 4b, we also show
the overlap of the GS $\Psi_0$ at $N$=16, $N_\Phi$=29 with a trial state
$|$Pair$>$, which is the GS at $V_1$=1.1 where the gap
is maximal. The rapid drop of the overlap $<$Pair$|\Psi_0$$>$,
as $V_1$ is reduced below one, very similar to the one observed for
$<$Pfaffian$|$GS$>$ at $N$=8 (cf. Fig.3b), is another indicator for
the phase transition to the compressible state at small $V_1$.
This transition is associated with a small wavevector instability in the
excitation spectrum. In our spherical system, it occurs
at $L$=2 both for $N$=8 and 16. This compressible state is
not a CF-liquid. It might be the charge density wave              \hfill
\begin{figure}
\center
\epsfxsize=3.2truein
\hskip 0.0truein \epsffile{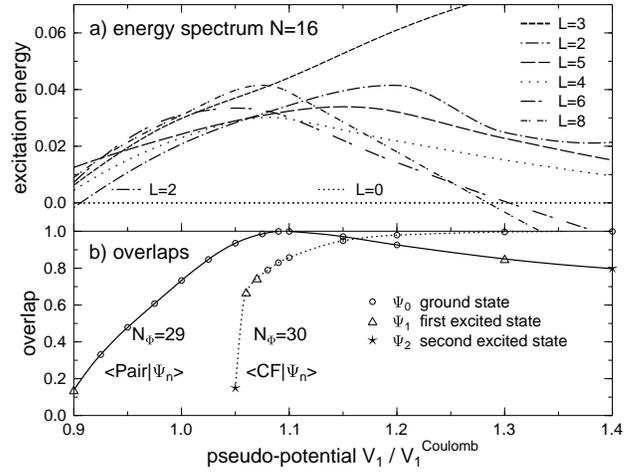}
\vspace{0.1in}
\caption{
a) Same as Fig. 3a but for $N$=16, $N_{\Phi}$=29.
This larger system becomes gapless too for small and large $V_1$.
b) Overlap of the GS $\Psi_0$ at $N$=16,$N_{\Phi}$=30 with the CF-liquid wf
vs. $V_1$ (dotted line). For large $V_1$, the overlap approaches unity.
Overlap of the GS $\Psi_0$ at $N$=16,$N_{\Phi}$=29 with the `trial' state
$|$Pair$>$, defined as GS for maximal gap (full line).
}
\label{fig:fig4}
\end{figure}
\noindent
state proposed by Koulakov et al. \cite{Koulakov}.
To study such states, the torus geometry may be more appropriate.
\begin{figure}
\center
\epsfxsize=3.2truein
\hskip 0.0truein \epsffile{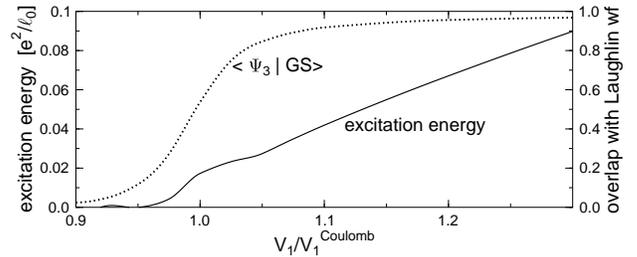}
\vspace{0.1in}
\caption{
Overlap of the GS with the Laughlin state $\Psi_3$ and
excitation energy of 10 electron system vs. $V_1$ at $\nu$=7/3.
}
\label{fig:fig5}
\end{figure}
It is instructive to study the system at the nearby $\nu$=7/3 filling since
Hall plateaux at 5/2 and 7/3 have been observed in the
same experiment \cite{Eisenstein90}. The results for energy gap and
GS overlap with the 1/3-Laughlin state $\Psi_3$ shown in Fig. 5, are
evidence for a phase transition from a gapless at small $V_1$ to an
incompressible state at around $V_1$$\approx$0.96. The energy gap
for Coulomb interaction, $V_1$=1, is $\Delta_{7/3}$$\approx$0.02
which is close to the calculated value at 5/2.
In the activation studies of Eisenstein et al. \cite{Eisenstein90}, it was
found that the gap at $\nu$=7/3 decreases with increasing tilt angle and
disappears in much the same way as at $\nu$=5/2 \cite{Eisenstein_private}.
As the FQH state at 7/3 is almost certainly spin-polarized,
and according to our numerical results at 5/2 likewise, a common origin for
the reduction of the gaps with increasing tilt angle and for their
disappearance may be expected. Our results imply that a reduction
of $V_1$ would simultaneously reduce both gaps and eventually lead to
compressible states. Besides increasing the Zeeman energy, which cannot account
for gap reduction in polarized states, a tilted $B$-field breaks                      \hfill
\begin{figure}
\center
\epsfxsize=3.2truein
\hskip 0.0truein \epsffile{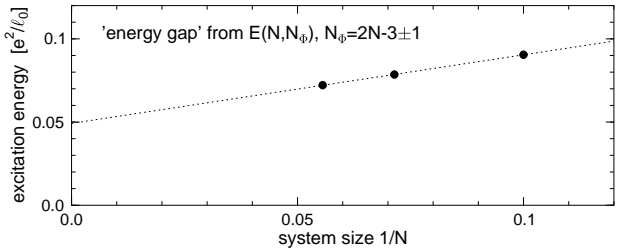}
\vspace{0.1in}
\caption{
Energy gap $\tilde{\Delta}$ of polarized system at $\nu_1$=1/2 vs.
system size $1/N$, calculated from charged excitations.
}
\label{fig:fig6}
\end{figure}
\noindent
rotational invariance
in the plane and leads to a coupling between in-plane and perpendicular
degrees of freedom, whose precise effect is not yet well understood.
If the main effect of the in-plane $B$-field was
a compression of the wf in the perpendicular direction,
as is often assumed \cite{Eisenstein_book}, an
increase of $V_1$ \cite{V1} and a corresponding enhancement of the gap
at 7/3 would result, in conflict with experiment.
But, if modifications to the electron interaction are indeed the cause for
the gap reduction and the transition to compressible states at both 5/2 and
7/3, as we believe, a reduction of $V_1$ by just a few percent would suffice
to explain the observed behavior.

Finally, in Fig. 6, we present the results of an alternative calculation of
the gap from the energy of charged excitations,
by changing the flux $N_\Phi$ by $\pm 1$, cf. \cite{dAmb}. 
Sizes other than those presented are `aliased' by different types of FQH states
and cannot be used for calculation of the gap \cite{dAmb}.
The bulk limit
$\tilde{\Delta}$=0.050, obtained by extrapolation in $1/N$, is about twice
the value $\Delta$ obtained above from neutral excitations.
This is consistent with predictions \cite{HR,Greiter9192}
and with a GS for $S$=2,4 with angular momentum $L$=$O(N^0)$, in which two
quasi-particles are far apart to minimize their energy. At fixed field $B$,
the gap $\Delta_{5/2}$=0.025 is about 1/4 the gap
$\Delta_{1/3}$=0.102 at $\nu$=1/3 \cite{dAmb}, while
at fixed density it is $\approx$11 times smaller.

To conclude, our results imply that the `$g$-factor' determined from
experiment \cite{Eisenstein90} is not related
to spin, but represents a correlation energy which should scale
with $\sqrt{n_S}$. A detailed study of activation energies
at $\nu$=5/2 and 7/3 for samples of different densities together with a
reliable calculation of tilted field effects will help decide if the
key piece in the $\nu$=5/2 puzzle has now been found.
But, the nature of the compressible state at small $V_1$ remains an open
question.

I thank J. Dousson and A. Possoz for speeding-up the
computer code, and F. Schlep\"utz for providing a
DEC-Alpha workstation with 1 Gigabyte of memory.

I am grateful for stimulating discussions with N. Cooper, D. Leadley,
E. Rezayi, B. Shklovskii, R. Willett and in particular with N. d'Ambrumenil,
J. Eisenstein, B.I. Halperin and H. St\"ormer who often encouraged me and
helped improve the manuscript.

\widetext

\begin{references}
\bibitem{HRprime}
If pairing of electrons plays a role,
we are guided by the arguments of ref. \cite{HR}: The flux of a paired state
at $\nu$=1/2 is $N_\Phi$=2$N$-4+$L_{rel}$, where $L_{rel}$ is
the relative angular momentum of the electrons in a pair.
Symmetry requires even (odd) $L_{rel}$ for spin-singlet (triplet) pairs.
For $L_{rel}$$>$2 there is no true pairing, as half-filling implies $L_{rel}$=2
in the mean. This leaves the possibilities $S$=2,4 for spin-singlet and
$S$=3 for polarized states.
\bibitem{Hard_core} For $N$$\le$10, we verified that for the
interaction, $V_0$=$V_2$=1, and $V_L$=0 otherwise, zero energy unpolarized GS,
with $g(0)$=0 in the second LL, exist if and only if $N_\Phi>3N/2$.
\end{references}
\end{document}